\documentclass[twocolumn,numberedappendix,iop]{openjournal}
\usepackage{lineno}
\usepackage{graphicx,amsmath,amssymb,amstext}
\usepackage{amsbsy,amsfonts,amsthm,color}
\usepackage[colorlinks,linkcolor=blue,citecolor=blue,urlcolor=blue]{hyperref}
\usepackage[utf8]{inputenc}
\usepackage{float}
\usepackage{xcolor}
\usepackage{ulem}
\usepackage[T1]{fontenc}
\usepackage[title]{appendix}
\usepackage{savesym}
\savesymbol{tablenum}
\usepackage{siunitx}
\restoresymbol{SIX}{tablenum}
\DeclareSIUnit\h{\text{$h$}}

\begin{document}
\title{Impact of inhomogeneous curvature on growth rate measurements from magnitude fluctuations\vspace{-4em}}

\author{A.~Nguyen$^{1,2,*}$}
\author{C.~Blake$^{1,2}$}
\author{H.~J.~Macpherson$^{3,4}$}
\thanks{$^*$E-mail: andnguyen@swin.edu.au}

\affiliation{$^1$ Centre for Astrophysics and Supercomputing, Swinburne University of Technology, P.O. Box 218, Hawthorn, VIC 3122, Australia}
\affiliation{$^2$ OzGrav: The ARC Centre of Excellence for Gravitational Wave Discovery}
\affiliation{$^3$ Kavli Institute for Cosmological Physics, The University of Chicago, 5640 South Ellis Avenue, Chicago, Illinois 60637, USA}
\affiliation{$^4$ NASA Einstein Fellow}

\begin{abstract}

Our interpretation of current cosmological observations rests on the assumptions of homogeneity and isotropy, leading to uniform background curvature and expansion characterised by the Friedmann-Lema\^itre-Robertson-Walker (FLRW) spacetime metric. However, the large-scale structure of the Universe is non-uniform in detail, inducing inhomogeneous curvature and scale factor variations. In this paper, we use numerical cosmological simulations generated in full General Relativity to study the impact of inhomogeneous spacetime on the magnitude fluctuations of distant objects, focusing on their use as a probe of the growth rate of cosmic structure. We quantify the distortions in the magnitude correlation spectrum as a function of angular scale and redshift, and use these distortions to infer the systematic offset in the growth rate measurement. We find that at $z \lesssim 0.2$, the systematic offset in growth rate measurements between the full numerical relativity and FLRW treatments is sub-dominant to the statistical error of current datasets, confirming that FLRW modelling is adequate for current low-redshift peculiar velocity experiments.  Future datasets extending to higher redshift may require theoretical models that additionally incorporate the contributions of gravitational lensing and inhomogeneous curvature.
\\[1em]
\end{abstract}

\maketitle

\section{Introduction}
\label{sec:intro}

Foundational to modern cosmology, the Friedmann-Lema\^itre-Robertson-Walker (FLRW) metric assumes a homogeneous and isotropic spacetime. Adaptation of the FLRW metric to our observable Universe combines a cosmological constant, $\Lambda$, and cold dark matter (CDM) in the $\Lambda$CDM model, which has successfully described the results of many observations including the Cosmic Microwave Background (CMB) anisotropies \citep[e.g.,][]{2020A&A...641A...6P}, standard candle observations of type Ia supernovae \citep[SNe; e.g.][]{2022ApJ...938..110B, 2024ApJ...973L..14D, 2025ApJ...986..231R}, standard ruler observations of the baryon acoustic peak \citep[e.g.,][]{2017MNRAS.470.2617A, 2025PhRvD.112h3515A} and weak gravitational lensing \citep[e.g.,][]{2022PhRvD.105b3520A, 2023PhRvD.108l3519D, 2025A&A...703A.158W}.

The $\Lambda$CDM model has provided a consistent fit to many observations, however, tensions between inferred parameters have arisen in some cases.  One example is the Hubble tension: a discrepancy between the value of the Hubble constant, $H_0$, inferred from CMB observations \citep{2020A&A...641A...6P}, compared to that determined by the cosmic distance ladder \citep[e.g.,][]{2021ApJ...919...16F, 2022ApJ...934L...7R, 2025ApJ...985..203F}. Explanations proposed for this tension include extensions to $\Lambda$CDM, dynamical dark energy, early dark energy, additional relativistic species, or modified gravity models \citep[for recent reviews, see][]{2021CQGra..38o3001D, 2022PhR...984....1S, 2025PDU....4901965D}.  An interesting area for investigation is the extent to which the assumption of the FLRW metric contributes to observable tensions.

The FLRW metric arises from the cosmological principle, which posits that on large enough scales, the Universe is homogeneous and isotropic. There is strong evidence, including CMB measurements, supporting that the Universe is statistically homogeneous and isotropic on large scales \citep{2020A&A...641A...6P}. The scale of homogeneity has been measured at around $70 - 80 \, h^{-1}$\unit{Mpc} \citep[e.g.][]{2005ApJ...624...54H, 2012MNRAS.425..116S, 2017JCAP...06..019N, 2026MNRAS.548ag771B, 2026arXiv260102886G}, although other studies claim that structure exists above these scales \citep[e.g.][]{2013MNRAS.429.2910C,2015A&A...584A..48H}.
There have been significant efforts to test homogeneity and isotropy, with many measurements in agreement with the cosmological principle, while some others claim evidence for deviations \citep[for a recent review, see][]{2023CQGra..40i4001A}.

Mapping the distribution of galaxy clusters and voids shows that the Universe is not homogeneous or isotropic on smaller scales. In non-linear General Relativity, the averaged time evolution of an inhomogeneous universe differs from the time evolution after spatial averaging \citep{1997A&A...320....1B,2000GReGr..32..105B,2001GReGr..33.1381B, 2007PhRvL..99y1101W}.
It has been proposed that such differences can potentially lead to ``backreaction'' effects, which may influence the large-scale expansion of the Universe \citep{2012ARNPS..62...57B}. It has long been debated whether these effects may provide a possible explanation for the observed large-scale acceleration of the Universe's expansion without invoking dark energy \citep[e.g.,][]{2011PhRvD..83h4020G, 2012PhRvD..85f3512G, 2013PhRvD..87l4037G, 2014CQGra..31w4003G, 2015CQGra..32u5021B, 2021FrASS...8..113T}.  Whether or not large-scale effects are possible,  as cosmological observations become increasingly precise, such effects may nonetheless have observable consequences on smaller scales, potentially sufficient to influence local-Universe observations \citep{2015PhRvL.114e1302A}. \citet{2019CQGra..36a4001A} and \citet{2019PhRvD..99f3522M} have used general relativistic simulations to provide additional evidence that these effects, while unlikely to entirely explain dark energy, may have relevance for precision cosmology.

Where previous studies have investigated the role of inhomogeneous cosmological models in addressing the Hubble tension \citep[e.g,][]{2023arXiv230512819P, 2024JCAP...01..071G, 2024JCAP...08..052C}, we focus on how these models affect our interpretation of another local-Universe observable: peculiar velocities (PVs) \citep[for a recent review of PV cosmology, see][]{2024arXiv241119484T}.  When interpreted in the context of a homogeneous model, PVs are the deviations in galaxy motion from the Hubble flow, induced gravitationally by the surrounding density field. They are typically deduced from simultaneous observations of redshift-independent standard-candle distances and observed redshifts of objects, and may be usefully quantified as a magnitude or luminosity distance fluctuation of a source at a given observed redshift \citep{hui2006, amendola2021}.  In such models, correlations between these PVs, or magnitude fluctuations,  are a powerful probe of the growth rate of structure, $f$, useful for discerning competing theories of gravity \citep[e.g.,][]{strauss1995, davis2011b, 2024arXiv241119484T, 2026OJAp....961766N}.

PV datasets have grown by an order of magnitude in recent years, and increases of a similar factor are anticipated from ongoing and upcoming surveys using distance indicators such as the Fundamental Plane, the Tully-Fisher relation, and SNe Ia \citep[examples of recent datasets and compilations are described by][]{2016AJ....152...50T, 2023ApJ...944...94T, 2025A&A...694A...1R, 2025arXiv251203226R, 2026ApJ..1001...20D}.  The current generation of PV surveys will soon reach close to full coverage of the sky to a redshift of $z = 0.1$ \citep{2023MNRAS.525.1106S, 2023Msngr.190...46T, 2023MNRAS.519.4589C}.

In the FLRW framework, a PV is defined with respect to the homogeneous background.  For some inhomogeneous models there is no such global background: the curvature, scale factor, and cosmic time vary as a function of position. As a result, the observed magnitude fluctuation of a source reflects not only its motion through the local density field, but also the effects of inhomogeneous curvature near the source as well as along the line of sight. Here, we use ``inhomogeneous curvature'' to encapsulate all effects beyond the standard modelling of magnitude fluctuations which incorporates FLRW and PVs. Constraints on the growth rate of structure obtained assuming a perturbed FLRW metric in the presence of inhomogeneities might therefore be systematically biased. Likewise, magnitude fluctuation correlations in current datasets can potentially be used to test inhomogeneous cosmological models, if the appropriate theoretical templates can be established \citep{2016NatPh..12..346A}.

Various frameworks are available for assessing the impact of inhomogeneous cosmological models on observables. For example, \citet{2013PhRvL.110x1305M} applied perturbations to the $\Lambda$CDM model, while \citet{2025PhRvL.135g1004G} parameterise inhomogeneities as non-virialised but non-linearly evolving overdense and underdense regions. Beyond-$\Lambda$CDM frameworks include the ``swiss cheese'' model \citep[e.g.][]{2007PhRvD..76l3004M, 2008JCAP...06..021B, 2014JCAP...06..054F, 2013JCAP...12..051L, 2025ApJ...994...50C}; the ``timescape'' model \citep[e.g.][]{2009PhRvD..80l3512W, 2011MNRAS.413..367S, 2013PhRvD..88h3529W}; the Lema\^itre-Tolman-Bondi (LTB) model \citep[e.g.][]{2006PhRvD..73h3519A, 2008PhRvL.101m1302C, 2011PhRvD..83f3506N, 2020MNRAS.491.2075L, 2022MNRAS.509.1291C, 2022CQGra..39r4001C}; and the Szekeres model \citep[e.g.][]{2016JCAP...06..035B, 2017JCAP...06..025B, 2024PhRvD.110l3526C, 2025arXiv251216591G}.

In our study we explore these phenomena using numerical simulations in non-linear General Relativity, which replace the enforced background of the FLRW metric with a self-consistent treatment of varying curvature and scale factor. Such simulations allow us to test the impact of inhomogeneous curvature on cosmological measurements. Examples of these tests in the literature include those using general relativistic N-body simulations \citep[e.g.][]{2015PhRvL.114e1302A, 2016JCAP...07..053A}. Numerical Relativity (NR) simulations have been used to study the growth of linear perturbations into the non-linear regime \citep{2016PhRvL.116y1302B, 2016PhRvL.116y1301G, 2017PhRvD..95f4028M, 2023PhRvD.108j3534M}, the backreaction of large-scale structure formation \citep{2018PhRvD..97d3509E, 2019PhRvD..99f3522M}, and the impact on distance measurements \citep{2016ApJ...833..247G, 2017PhRvD..96j3530G, 2021PhRvD.104b3525M, 2023JCAP...03..019M, 2024ApJ...970..111M}.  In this work, we use NR simulations to investigate a currently unexplored aspect of cosmological backreaction: the role of inhomogeneous curvature as a source of systematic modelling error in measurement of the growth rate of structure using magnitude fluctuation correlations.

Our paper is structured as follows: in Section~\ref{sec:flrw} we outline how PVs are measured and modelled in the standard FLRW framework, and in Section~\ref{sec:simulations} we provide an overview of our NR simulations of non-linear structure formation. In Section~\ref{sec:methods} we describe how we create the magnitude fluctuation dataset and measure its correlation spectrum.  In Section~\ref{sec:results}, we compare the correlation measurements of the NR simulations with the FLRW baseline, and we discuss the implications of these results in Section~\ref{sec:discussion}. Finally, in Section~\ref{sec:conclusion}, we summarise our findings and suggest directions for future research.

\section{FLRW Cosmology}
\label{sec:flrw}

\subsection{FLRW metric}

The FLRW metric assumes an exactly homogeneous and isotropic spacetime. For a spatially flat universe the spacetime interval is given by,
\begin{equation}
    ds^2 = -c^2 \, dt^2 + a^2(t)\left( dx^2 + dy^2 + dz^2 \right),
    \label{eq:flrw}
\end{equation}
where $c$ is the speed of light, $t$ is cosmic time, $a(t)$ is the scale factor, and $x, y, z$ are the co-moving Cartesian coordinates. Cosmic expansion is given by the Friedmann equation, which in the flat $\Lambda$CDM model can be written as,
\begin{equation}
    H^2(z) = H_0^2 \left[ \Omega_m (1+z)^3 + \Omega_\Lambda \right],
    \label{eq:friedmann}
\end{equation}
where $H(z) = \dot{a}/a$ is the Hubble parameter, $\dot{a}$ is the cosmic time derivative of $a(t)$ also called the expansion velocity, $\Omega_m$ and $\Omega_\Lambda$ are the present-day matter and dark energy density parameters, $H_0 = H(z=0)$ is the Hubble constant, and the scale factor has been parameterised by redshift $z$, where $a=1/(1+z)$.

In this cosmological model, the radial co-moving coordinate of a source at redshift $z$ is,
\begin{equation}
    \chi(z) = \int_0^z \frac{c \, dz'}{H(z')},
\end{equation}
and in FLRW specifically, the observable angular diameter and luminosity distances are given by $D_A = \chi(z) / (1+z)$ and $D_L = (1+z) \, \chi(z)$. These are related by the Etherington reciprocity relation,
\begin{equation}
    D_L = D_A \, (1+z)^2 ,
    \label{eq:erthington}
\end{equation}
which holds true in any metric theory of gravity.

\subsection{Peculiar velocities}

An FLRW universe is modelled by the homogeneous background expansion (``Hubble flow'') generating cosmological source redshifts, $z_{\rm cos}$. Deviations from the Hubble flow, known as peculiar velocities within $\Lambda$CDM, arise gravitationally from the matter density field, such that the observed redshift of a source is given by,
\begin{equation}
    1+z \approx (1+z_{\rm cos})(1+v_r/c) ,
\end{equation}
where $v_r$ is the radial PV component.  In linear perturbation theory, the three-dimensional PV of a source at position $\mathbf{x}$ can be related to the matter density contrast $\delta(\mathbf{x})$ through the linearised continuity equation,
\begin{equation}
    \mathbf{v}(\mathbf{x}) = \frac{a H f}{4\pi} \int d^3 \mathbf{x}' \, \frac{\delta(\mathbf{x}') \, (\mathbf{x}' - \mathbf{x})}{|\mathbf{x}' - \mathbf{x}|^3} ,
    \label{eq:pecvel}
\end{equation}
where $f \approx \Omega_m^{0.55}$ is the linear growth rate of structure in $\Lambda$CDM cosmology \citep{2005PhRvD..72d3529L}.  PVs are hence an important cosmological probe because they are directly sensitive to the growth rate of structure, which encodes information about gravity and the matter content of the Universe.

In the FLRW model, PVs are inferred from the deviation of a source's observed luminosity distance from the value predicted by the Hubble flow at its observed redshift; hence, these measurements require standard-candle distance indicators. In general cosmological models, this deviation in luminosity distance may be equivalently characterised by a magnitude fluctuation, $\delta m$. In a perturbed universe, $\delta m$ is related to the observed magnitude $m$ and the magnitude expected at the observed redshift in a non-perturbed homogeneous universe, $\overline{m}$, by $\delta m = m - \overline{m}$. For a given $D_L$, $\delta m$ is given by,
\begin{equation}
    \delta m = \frac{5} {\ln{10}} \frac{\delta D_L}{D_L},
    \label{eq:dm}
\end{equation}
where the luminosity distance fluctuation is given by $\delta D_L = D_L - \overline{D}_L$, where $D_L$ is the observed luminosity distance and $\overline{D}_L$ is the luminosity distance at the observed redshift in a homogeneous universe.

For a source with radial PV $v_{r,\rm src}$ and an observer with radial velocity $v_{r,\rm obs}$, the predicted magnitude fluctuation, $\delta m_{\rm pred}$, is given by,
\begin{equation}
    \delta m_{\rm pred} = \alpha \left( v_{r,\rm src} - v_{r,\rm obs} \right),
    \label{eq:dmpred}
\end{equation}
where,
\begin{equation}
    \alpha = \frac{5}{\ln 10} \cdot \frac{1}{c} \left[ 1 - \frac{c \, (1+z)^2}{H(z) \, D_L} \right]
    \label{eq:alpha}
\end{equation}
encodes the sensitivity of apparent magnitude to PV
\citep{hui2006, amendola2021}.

\subsection{Magnitude correlation spectrum due to velocities}
\label{sec:cl_velocity}

Given that PVs and magnitude fluctuations are induced by the underlying density fluctuations of large-scale structure, they contain a correlation pattern which depends on the underlying matter power spectrum, $P(k)$, as a function of wavenumber $k$.  In this paper, we quantify the velocity and magnitude correlations at a given observed redshift using their angular power spectra, $C_\ell^{v_r \, v_r}$ and $C_\ell^{\delta m \, \delta m}$, as a function of multipole $\ell$.

The theoretical angular power spectrum of velocities in the FLRW model is given by,
\begin{equation}
    C_\ell^{v_r v_r} = \frac{2}{\pi} \int_0^\infty dk \, k^2 \, P(k) \, \left[ W_\ell^v(k) \right]^2 ,
\label{eq:cluu}
\end{equation}
where the velocity window function $W_\ell^v(k)$ follows from a recursion relation,
\begin{equation}
    (2\ell + 1) W^v_\ell(k) = \frac{H f a}{k} \left[ \ell \, W^\delta_{\ell-1}(k) - (\ell + 1) W^\delta_{\ell+1}(k) \right] ,
    \label{eq:wlu}
\end{equation}
in terms of the density window function,
\begin{equation}
    W^\delta_\ell(k) = \int_{\chi_{\rm min}}^{\chi_{\rm max}} d\chi \, w(\chi) \, j_\ell(k\chi) ,
    \label{eq:wld}
\end{equation}
where $j_\ell$ is a spherical Bessel function of order $\ell$, and $w(\chi)$ is the distance distribution of the sources, normalised such that $\int_{\chi_{\rm min}}^{\chi_{\rm max}} d\chi \, w(\chi) = 1$, where $\chi_{\rm min}$ and $\chi_{\rm max}$ are the lower and upper distance boundaries of the source distribution.  We present a derivation of these relations in Appendix \ref{sec:derivation}.

In this work, we consider a single observed spherical shell of sources at a given observed redshift (and corresponding distance $\chi_0$).  The radial selection function is hence a delta-function $w(\chi) = \delta_D(\chi - \chi_0)$ and we find,
\begin{equation}
    \begin{split}
        W^v_\ell(k) &= \frac{Hfa}{k} \, j'_\ell(k \chi_0) , \\
        C^{v_r v_r}_\ell &= \frac{2 \, H^2 f^2 a^2}{\pi} \int dk \, P(k) \left[ j'_\ell(k \chi_0) \right]^2 . \\
    \end{split}
\end{equation} 
Since the magnitude fluctuation in Eq.~\eqref{eq:dmpred} is proportional to the radial PV, the predicted FLRW angular power spectrum of magnitude fluctuations due to velocities is hence,
\begin{equation}
    C_\ell^{\delta m \, \delta m}\big|_{\rm FLRW velocities} = \alpha^2 \, C_\ell^{v_r v_r}.
    \label{eq:cldm_velocity}
\end{equation}
Since $C_\ell^{v_r v_r} \propto f^2$, the magnitude fluctuation power spectrum is hence a direct probe of the growth rate of structure.  We used \texttt{CAMB}\footnote{\url{https://camb.info}} \citep{lewis2000} to compute the linear matter power spectrum $P(k)$ matching the initial conditions of the NR simulation, as described in Sec.~\ref{sec:simulations} below.

\subsection{Magnitude correlation spectrum due to lensing}
\label{sec:cl_lensing}

The correlation spectrum of magnitude fluctuations also contains the imprint of weak gravitational lensing, whose effect is described using the lensing convergence, $\kappa = - \delta D_L / \overline{D}_L$ \citep[e.g.,][]{2001PhR...340..291B, 2015RPPh...78h6901K, 2018ARA&A..56..393M}.  The theoretical angular power spectrum of the lensing convergence in the FLRW model is given by,
\begin{equation}
C_\ell^{\kappa\kappa} = \left( \frac{3 \Omega_m H_0^2}{2 c^2} \right)^2 \int_0^{\chi_H} d\chi \, \frac{[W^\kappa(\chi)]^2}{a^2(\chi)} P\left(\frac{\ell}{\chi}, \chi \right) ,
\end{equation}
where $\chi_H$ is the horizon distance, and $W^\kappa(\chi)$ is the convergence window function of the sources depending on their distance distribution $w(\chi)$ defined above,
\begin{equation}
W^\kappa(\chi) = \int_\chi^{\chi_H} d\chi_s \, w(\chi_s) \left( \frac{\chi_s - \chi}{\chi_s} \right) .
\end{equation}
Assuming a delta-function source distribution as in Sec.~\ref{sec:cl_velocity}, and an Einstein de-Sitter growth history appropriate to our simulations described in Sec.~\ref{sec:simulations}, such that $P(k,\chi) = P(k,0) \, a^2(\chi)$, we find,
\begin{equation}
C_\ell^{\kappa\kappa} = \left( \frac{3 \Omega_m H_0^2}{2 c^2} \right)^2 \int_0^{\chi_0} d\chi \left( \frac{\chi_0 - \chi}{\chi_0} \right)^2 P\left(\frac{\ell}{\chi}, 0 \right) .
\end{equation}
Using Eq.~\eqref{eq:dm}, the predicted FLRW angular power spectrum of magnitude fluctuations due to lensing is then,
\begin{equation}
    C_\ell^{\delta m \, \delta m}\big|_{\rm FLRW \; lensing} = \left( \frac{5}{\ln{10}} \right)^2 C_\ell^{\kappa\kappa}.
\label{eq:cldm_lensing}
\end{equation}
We will compare these predictions to the measurements from the simulations described below, in order to assess the importance of inhomogeneous curvature effects.

\section{Numerical Relativity Simulations}
\label{sec:simulations}

\subsection{Inhomogeneous curvature}

In an inhomogeneous universe, the spatial curvature is not constant as it is in FLRW, but varies as a function of position as a consequence of the non-uniform distribution of matter. The averaged evolution of an inhomogeneous spacetime in general differs from that of FLRW, leading to additional terms in the effective Friedmann equation called ``backreaction'' terms \citep[e.g.,][]{1997A&A...320....1B,2000GReGr..32..105B,2001GReGr..33.1381B}.  These inhomogeneous curvature terms also affect the propagation of light through the inhomogeneous spacetime. As photons traverse regions of varying gravitational potential, their geodesics are altered compared to the FLRW prediction \citep{2006PhRvD..73b3523B}. The angular diameter distance, $D_A$, and redshift, $z$, both acquire direction-dependent corrections that depend on the observer position and local curvature along the line of sight.

In NR, the Einstein field equations are solved directly on a computational grid. The line element takes the Arnowitt-Deser-Misner form \citep{1959PhRv..116.1322A},
\begin{equation}
    ds^2 = -\alpha_{\rm{NR}}^2 \, c^2 \, dt_{\rm c}^2 + \gamma_{ij} \left( dx^i + \beta^i \, dt_{\rm c} \right) \left( dx^j + \beta^j \, dt_{\rm c} \right),
    \label{eq:adm}
\end{equation}
where $\alpha_{\rm NR}$ is the lapse function, $t_{\rm c}$ is the coordinate time, $\beta^i$ is the shift vector, and $\gamma_{ij}$ is the spatial metric on each hypersurface.

\begin{figure*}
    \centering
    \includegraphics[width=\textwidth]{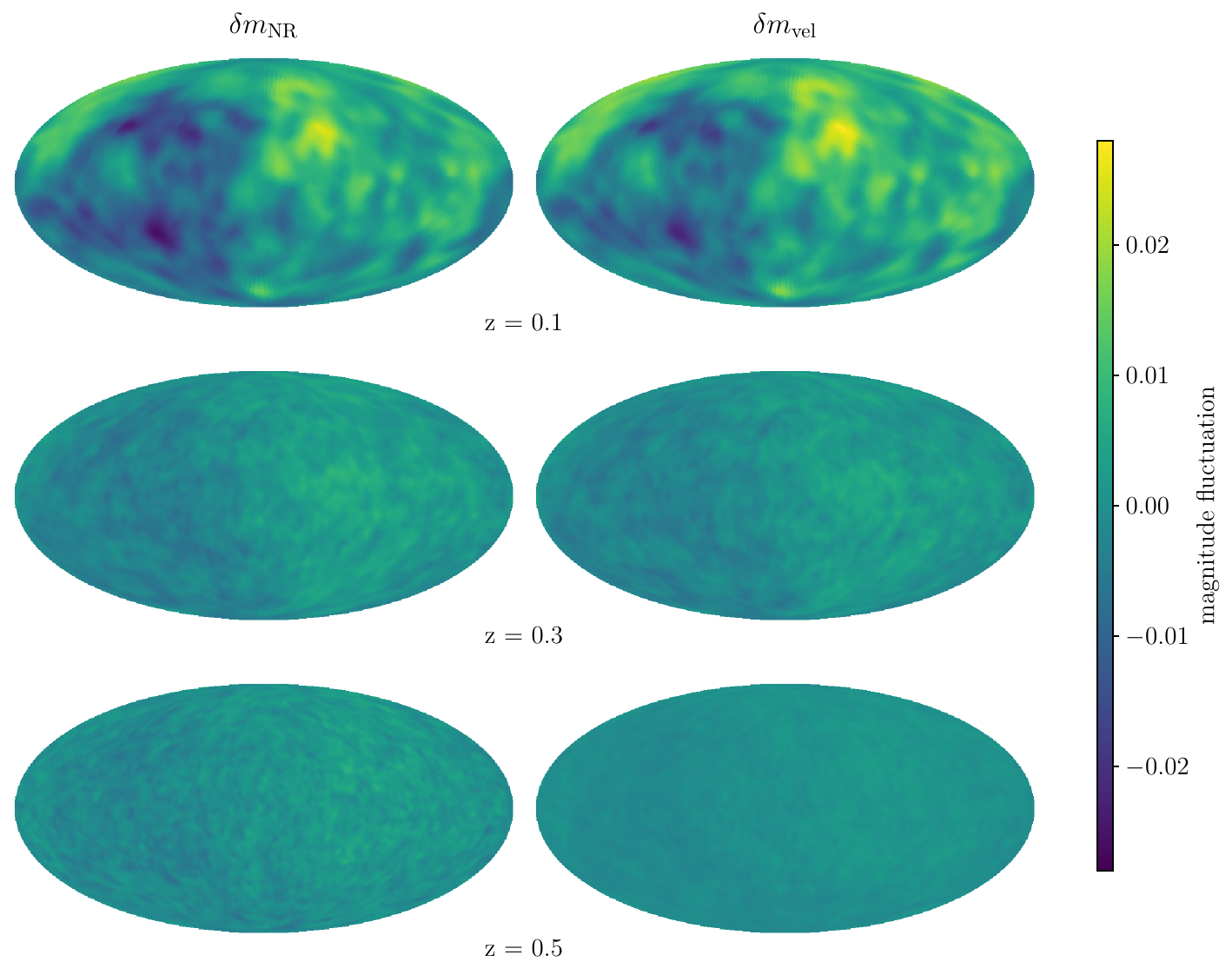}
    \caption{Projections of the magnitude fluctuation field $\delta m$ at redshifts $z = 0.1$, $0.3$, and $0.5$ (top to bottom). The left column shows the full NR measurement $\delta m_{\rm NR}$, and the right column shows the velocity-only values $\delta m_{\rm vel}$ calculated using the source and observer velocities. At $z = 0.1$ the two maps are visually similar, with large-scale features driven dominantly by velocities. At higher redshifts, the amplitude of the fluctuations decreases and the maps become increasingly distinct, with smaller scale structure formation apparent in $\delta m_{\rm NR}$ and absent from the $\delta m_{\rm vel}$ maps, reflecting the growing contribution of gravitational lensing to the signal.}
    \label{fig:magfluc}
\end{figure*}

\subsection{Simulations}
\label{sec:nrsim}

We use the \texttt{Einstein Toolkit}\footnote{\url{https://www.einsteintoolkit.org}} \citep[ET;][]{Zilhao:2013,Loffler:2012}, which is based on the Cactus\footnote{\url{https://www.cactuscode.org}} infrastructure, for our NR simulations. Cosmological initial data is generated for the ET via \texttt{FLRWSolver} \citep{2017PhRvD..95f4028M}, 
for which we specify the \texttt{CAMB} $P(k)$ at the initial redshift $z_{\rm ini}=20$. On the initial slice of the simulations, we describe the metric as a perturbed FLRW metric in Newtonian gauge. We emphasise that this is an assumption on the initial slice only, and the evolution does not explicitly enforce any FLRW-like evolution. The fluctuations on the initial slice are constructed so that at redshift $z=0$ the matter power spectrum roughly coincides with the observed $\Lambda$CDM $P(k)$ on the sampled scales, where we assume $\Lambda$CDM parameters $A_s = 2 \times 10^{-9}$, $k_{\rm pivot} = 0.05 \, h/{\rm Mpc}$, $n_s = 0.965$, Hubble constant $h = 0.7$ (where $H_0 = 100 h$ km/s/Mpc), and baryonic and CDM energy content $\Omega_b = 0.049$ and $\Omega_{\rm cdm} = 0.25 - \Omega_b$.

Our fiducial simulation has resolution $N=256$ (total cubic grid with $N^3$ cells) and box length $L=3072\,h^{-1}$ Mpc. We use two lower-resolution simulations with $N=128$ and 200, which sample the same physical scales, to ensure our results are robust to numerical resolution (see Fig.~\ref{fig:clallres}). The largest mode we sample is set by the size of the entire domain, and nominally the smallest mode is set by the grid resolution. However, we remove modes with wavelength less than 10 grid cells from the initial data, amounting to a minimum scale of $\sim 120\,h^{-1}$ Mpc on the initial slice. This reduces the amount of numerical error in the simulation due to under-sampling of small-scale modes. While structure below this scale does grow as the simulation becomes non-linear, it is highly damped with respect to the expected power on these scales.  We incorporate this effect in the theoretical modelling described in Sec.~\ref{sec:cl_velocity} and Sec.~\ref{sec:cl_lensing} by setting the power spectrum amplitude to zero for $k > 0.09 \, h$ Mpc$^{-1}$, corresponding to the scale below which we remove modes.

Lastly, the simulations do not contain a cosmological constant, $\Lambda$. This is simply because the evolution of Einstein's equations incorporating a cosmological constant has not yet been included in the ET.  Further specifics on the thorns (modules within the Cactus infrastructure) used in the evolution, the gauge choices we make, and any details of the simulations not mentioned here can be found in \citet{2026arXiv260324838M}, who use the same simulations as we do.

Once the simulations are completed, we use \texttt{mescaline}\footnote{Only the spatial averaging part of \texttt{mescaline} is publicly available: \url{https://github.com/hayleyjm/mescaline-1.0} \citep[as used in, e.g.][]{2019PhRvD..99f3522M}. Ray tracing will be added to the public code in the near future.} \citep{2019PhRvD..99f3522M} to perform general-relativistic ray tracing. Like our simulations, \texttt{mescaline} also does not make any physical assumptions on the form of the metric tensor and is completely general (for zero shift vector). See \citet{2023JCAP...03..019M} for details and tests of the \texttt{mescaline} ray tracer. 

We place a family of 20 synthetic observers at random locations on the simulation hypersurface with $z=0$, and propagate an isotropic set of geodesics back in time through the simulation output. The resulting outputs are light-cone slices with constant simulation time for each observer. 
\texttt{Mescaline} outputs redshifts and angular diameter distances in the frame co-moving with the fluid flow in the simulation.

The initial direction of geodesics are set as the centre of \texttt{HEALPix}\footnote{\url{https://HEALPix.sourceforge.io}} \citep{Gorski:2005} pixels with $N_{\rm side}=32$; resulting in 12,288 individual lines of sight for each observer. We note that the deflection of the geodesic is calculated exactly by \texttt{mescaline}, and the \texttt{HEALPix} resolution only impacts calculations such as angular power spectra when considering full-sky maps.

\begin{figure*}
    \centering
    \includegraphics[width=\textwidth]{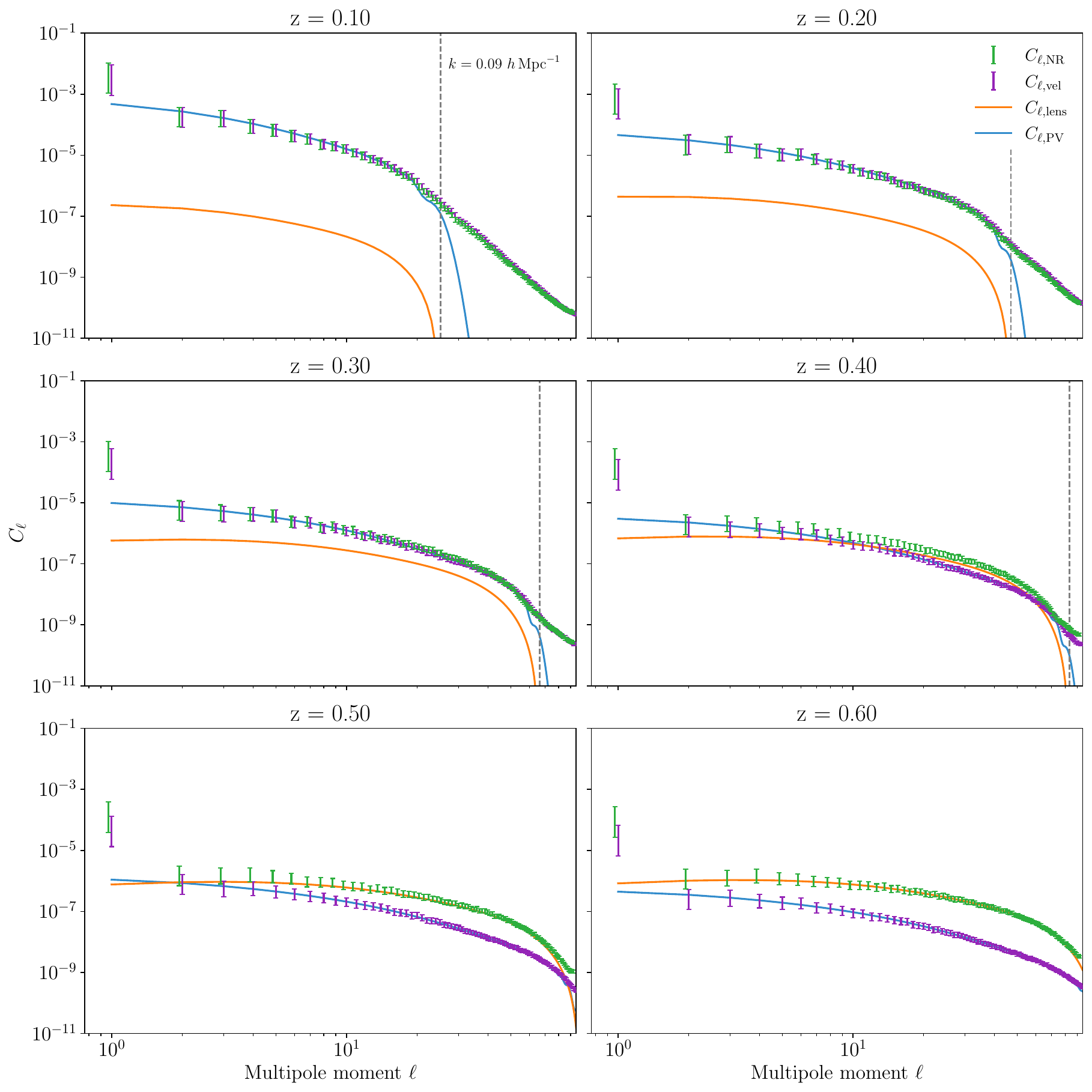}
    \caption{The angular power spectrum of magnitude fluctuations, $C_\ell^{\delta m \, \delta m}$, at a range of redshifts from $z = 0.1$ to $z = 0.6$. The green points show the full NR simulation measurement ($C_{\ell,{\rm NR}}$), the purple points show the velocity-only measurements ($C_{\ell,{\rm vel}}$), and the blue and orange curves show the FLRW model prediction due to velocities and lensing, respectively. The errors in the measurements are determined as the standard deviation across 20 observer positions. At low redshift, the full NR and velocity-only measurements agree closely. At higher redshift, the NR power spectra exceeds the velocity-only measurements due to the effects of gravitational lensing, with the excess appearing first at high $\ell$ and extending to all scales by $z \gtrsim 0.6$.  The dipole signature due to observer motion is visible at $\ell = 1$. The vertical dashed line shows a multipole $\ell \sim k \, \chi(z)$ corresponding to a constant physical wavenumber $k = 0.09 \, h$ Mpc$^{-1}$, which represents the power spectrum cutoff in the initial data.}
    \label{fig:cl256}
\end{figure*}

\section{Methodology}
\label{sec:methods}
\subsection{Extracting magnitude fluctuations}

From the ray-tracing outputs described in Sec.~\ref{sec:nrsim}, we constructed magnitude fluctuation maps at fixed observed redshift slices within the \texttt{HEALPix} pixelisation. The luminosity distance along each line of sight is obtained from the angular diameter distance reported by the simulation outputs via the reciprocity relation (Eq.~\eqref{eq:erthington}). The magnitude fluctuation at each pixel, $\delta m_{\rm NR}$, is then determined from the luminosity distance fluctuation at a fixed observed redshift using Eq.~\eqref{eq:dm}.  This quantity captures the full effect of the inhomogeneous spacetime on the observed brightness, including contributions from PVs, gravitational lensing, and inhomogeneous curvature. Since light rays within different pixels reach different redshifts at a given simulation time-slice, we perform a linear interpolation of $D_L$ and $z$ onto a set of observed redshifts $z \in [0.1, 0.6]$, spanning a range of current and future observational datasets. Magnitude fluctuation maps for an example observer, calculated from the NR simulations, are shown in the left-hand panels of Fig.~\ref{fig:magfluc}, with the top row at $z=0.1$, the middle row at $z=0.3$, and the bottom row at $z=0.5$.

\subsection{Constructing the velocity-only measurement}

We construct a companion ``velocity-only'' magnitude fluctuation, $\delta m_{\rm vel}$, using only the velocity information from the simulation, in order to isolate gravitational lensing or other inhomogeneous curvature effects.  We note that, whilst the velocity defined by simulation co-ordinates is not identical to the PV relative to a homogeneous background, the simulation frame is well-described on average by FLRW, so we do not expect these slight differences in definition to be important for our results.  The predicted value of $\delta m_{\rm vel}$ is computed via Eq.~\eqref{eq:dmpred}, using the velocities with respect to the simulation coordinates of the source and observer. This prediction approximately captures the Doppler contribution to $\delta m$ that would be present in a homogeneous FLRW universe with the same velocity field, but excludes the additional effects of inhomogeneous curvature and lensing. The velocity-only magnitude fluctuation maps calculated for an example observer are shown in the right-hand panels of Fig.~\ref{fig:magfluc}.

\subsection{Angular power spectrum estimation}

We estimated the angular power spectrum of magnitude fluctuations from the \texttt{HEALPix} maps using spherical harmonic decomposition. The magnitude fluctuation field on the sphere is expanded as,
\begin{equation}
    \delta m(\theta, \phi) = \sum_{\ell=0}^{\ell_{\rm max}} \sum_{m=-\ell}^{\ell} a_{\ell m} \, Y_{\ell m}(\theta, \phi),
\end{equation}
where $Y_{\ell m}$ are the spherical harmonics and $a_{\ell m}$ are the harmonic coefficients. We use the \texttt{Python} wrapper for \texttt{HEALPix}, \texttt{healpy}\footnote{\url{https://healpy.readthedocs.io}}, to calculate the coefficients via \texttt{healpy.anafast}. The ray tracing produces a \texttt{HEALPix} map with output $n_{\rm side} = 32$, and hence we measure multipoles up to the resolution limit $\ell_{\rm max} = 3 n_{\rm side} - 1 = 95$. The measured angular power spectrum is then given by,
\begin{equation}
    C_\ell^{\delta m \, \delta m} = \frac{1}{2\ell+1} \sum_{m=-\ell}^{\ell} |a_{\ell m}|^2 .
    \label{eq:cl_est}
\end{equation}
We compute $C_\ell$ separately for $\delta m_{\rm NR}$ and $\delta m_{\rm vel}$ at each redshift for individual observers, and subsequently average across all 20 observers.


\subsection{Fisher forecast for the growth rate offset}
\label{sec:fisher}

\citet{2008MNRAS.391..228A} provide a method for forecasting the statistical error and systematic bias for the best-fit parameters relative to a fiducial model, given an error vector and model distortion. We apply this to the inference of the growth rate $f$ that would result from using the velocity-only template to interpret a full non-linear solution. Assuming a simple 1-parameter cosmological model in terms of the growth rate, which is correct in the large-scale limit relevant for our analysis, the forecast statistical error in the growth rate is,
\begin{equation}
    \sigma_f = \frac{1}{\sqrt{F}} ,
\end{equation}
and the systematic offset in the growth rate is,
\begin{equation}
    \Delta f = \frac{B}{F} ,
\end{equation}
where $F$ is the Fisher matrix (just a scalar in this case, since we only have one parameter), and $B$ is the bias.  These variables are evaluated as,
\begin{equation}
    F = \sum_\ell \frac{1}{\sigma^2(C_\ell)} \left( \frac{dC_\ell}{df} \right)^2 ,
\end{equation}
and,
\begin{equation}
    B = \sum_\ell \frac{1}{\sigma^2(C_\ell)} \, \Delta C_\ell \, \frac{dC_\ell}{df} ,
\end{equation}
where $\sigma(C_\ell)$ is the error in the measurement, and $\Delta C_\ell$ is the power spectrum offset between the velocity-only measurement $C_{\ell,{\rm vel}}$ and the NR version, $C_{\ell,{\rm NR}}$,
\begin{equation}
    \Delta C_\ell = C_{\ell,{\rm NR}} - C_{\ell,{\rm vel}} .
\end{equation}
Since $C_\ell \propto f^2$, we can write $C_\ell = (f/f_{\rm fid})^2 C_{\ell,{\rm fid}}$, hence
\begin{equation}
    \frac{dC_\ell}{df} \big|_{f=f_{\rm fid}} = \frac{2f}{f_{\rm fid}^2} C_{\ell,{\rm fid}} = \frac{2}{f_{\rm fid}} C_{\ell,{\rm vel}} .
\end{equation}
Finally, the forecast error for an all-sky survey is,
\begin{equation}
    \sigma^2(C_\ell) = \frac{\left( C_\ell + N \right)^2} {2 \ell + 1},
\end{equation}
where the signal $C_\ell = C_{\ell,{\rm vel}}$, and the noise $N$ is,
\begin{equation}
    N = \frac{\sigma_m^2}{\sigma_{\rm source}} .
\end{equation}
In this expression, $\sigma_m$ is the noise in the magnitude measurement of each source, which is related to the accuracy of the standard candle, and $\sigma_{\rm source}$ is the angular density of the dataset in units of per steradian.  We chose $\sigma_m = 0.127$, which is characteristic of SNe Ia magnitude measurement noise in current surveys \citep[e.g.,][]{rigault2020, nicolas2021}.  We also assume an angular density $\sigma_{\rm source} = 10^4$ per steradian, characteristic of near-future surveys \citep[e.g][]{2023PASP..135j5002H, 2025A&A...694A...1R}.

\begin{figure*}
    \centering
    \includegraphics[width=\textwidth]{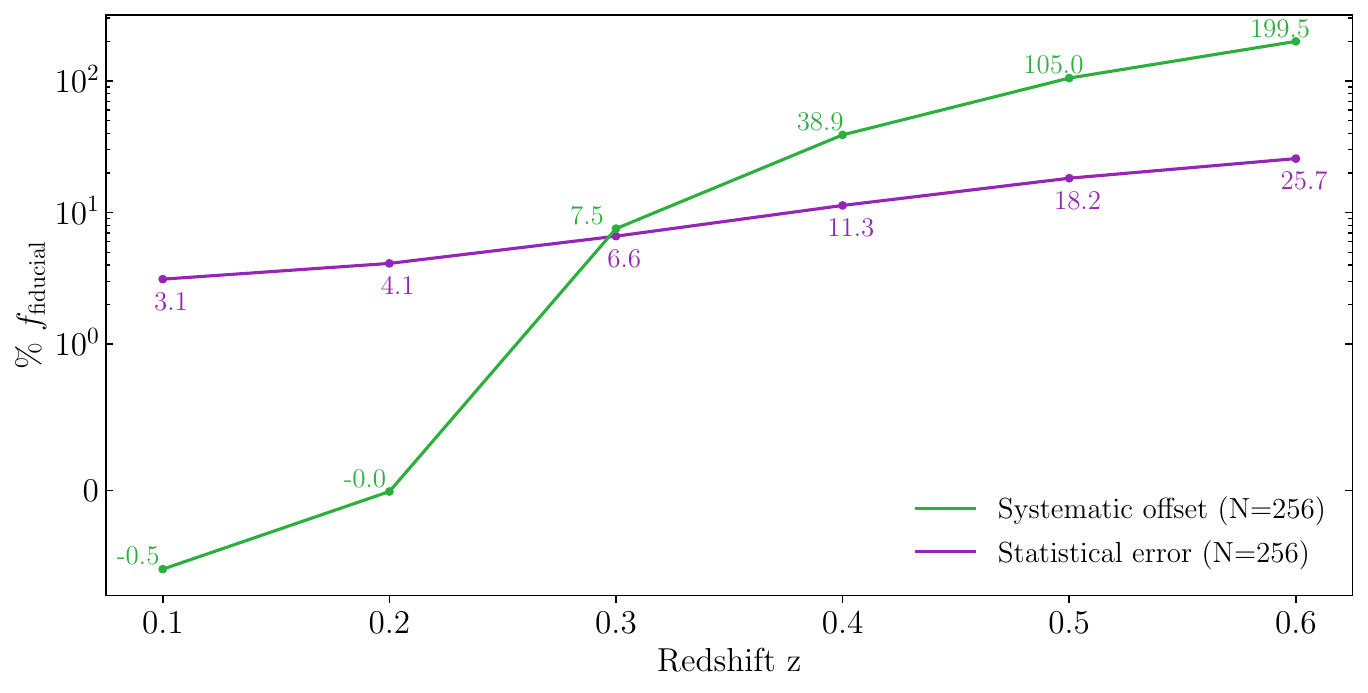}
    \caption{Fisher forecast of the systematic offset $\Delta f$ (green) and statistical error $\sigma_f$ (purple) in the growth rate, expressed as a percentage of the fiducial value $f_{\rm fid}$, as a function of redshift for the $N = 256$ resolution simulation. At $z \leq 0.2$ the systematic offset is negligible ($< 1\%$) compared to the statistical error ($3$--$4\%$). The systematic offset grows rapidly with redshift, reaching ${\sim} 8\%$ at $z = 0.3$, ${\sim} 39\%$ at $z = 0.4$, and exceeding $100\%$ by $z = 0.5$, while the statistical error increases more gradually. The crossover at $z \approx 0.3$ indicates the redshift beyond which curvature gradient and lensing effects would dominate over statistical uncertainties in growth rate measurements, for this future survey configuration.}
    \label{fig:sysoff256}
\end{figure*}

\section{Results}
\label{sec:results}

\subsection{Angular power spectra}
\label{sec:results_cl}

Our magnitude angular power spectrum measurements, for the NR simulation and the velocity-only case, are displayed in Fig.~\ref{fig:cl256} for redshifts in the range $z = 0.1$ to $z = 0.6$.  For comparison, we also show theoretical FLRW predictions for the effects of PVs (Eq.~\eqref{eq:cldm_velocity}) and lensing (Eq.~\eqref{eq:cldm_lensing}).  At each redshift, simulated angular power spectra are averaged over all 20 independent observer positions, with the error bars determined as the standard deviation across observers.  We note the presence of the dipole signal at $\ell=1$ is determined by the observer motion, which is degenerate with the growth rate, and so we exclude $\ell=1$ from further analysis.

At $z \lesssim 0.3$, the theoretical signal from lensing is negligible, and the full-simulation and velocity-only measurements are in close agreement with an average difference of $4.7\%$ for all $\ell > 1$. Photons from lower-redshift sources travel relatively short path lengths through the inhomogeneous spacetime, and so the accumulated effect of inhomogeneous curvature and gravitational lensing on the luminosity distance fluctuations is small. The magnitude fluctuations at low redshift are therefore expected to be well-captured by the FLRW prediction.

The theoretical velocity angular power spectrum shows good agreement with the measured velocity-only signal at all redshifts, up to a multipole $\ell \sim k \, \chi(z)$ corresponding to a constant physical wavenumber $k = 0.09 \, h$ Mpc$^{-1}$, which represents the power spectrum cutoff in the initial data, as discussed in Sec.~\ref{sec:nrsim}. These findings are consistent with previous studies based on perturbation theory and Newtonian simulations, which have explored the range of linear theory in PV analysis \citep{2006PhRvD..73b3523B, 2008PhRvD..78l3530B, 2014MNRAS.443.1900B, 2014MNRAS.445.4267K}. This agreement also validates the large-scale density fluctuation spectrum of the NR simulation, showing that the velocity field produces magnitude fluctuations consistent with linear perturbation theory predictions in this case. At higher $\ell$, the theoretical prediction falls below the velocity-only measurement, reflecting the fact that the simulation evolves some structure at $k > 0.09 \, h$ Mpc$^{-1}$ beyond the initial power spectrum cut-off scale, which is not captured by our modelling.

As redshift increases beyond $z \approx 0.3$, an excess in the NR angular power spectrum relative to the velocity-only measurement becomes apparent. This excess grows with increasing redshift. Across $0.3 \leq z \leq 0.6$ there is an average difference of $625 \%$ between $C_{\ell,{\rm NR}}$ and $C_{\ell,{\rm vel}}$ for all $\ell > 1$.  This increased power is well-matched by the theoretical lensing power spectrum, illustrating that at higher redshifts, the coherent magnification from gravitational lensing dominates the signal.  These comparisons also demonstrate that FLRW + PVs + lensing produces a good description of the magnitude fluctuations in the NR simulation, such that the residual inhomogeneous curvature effects are relatively small on the scales under consideration.

\begin{figure*}
    \centering
    \includegraphics[width=0.9\textwidth]{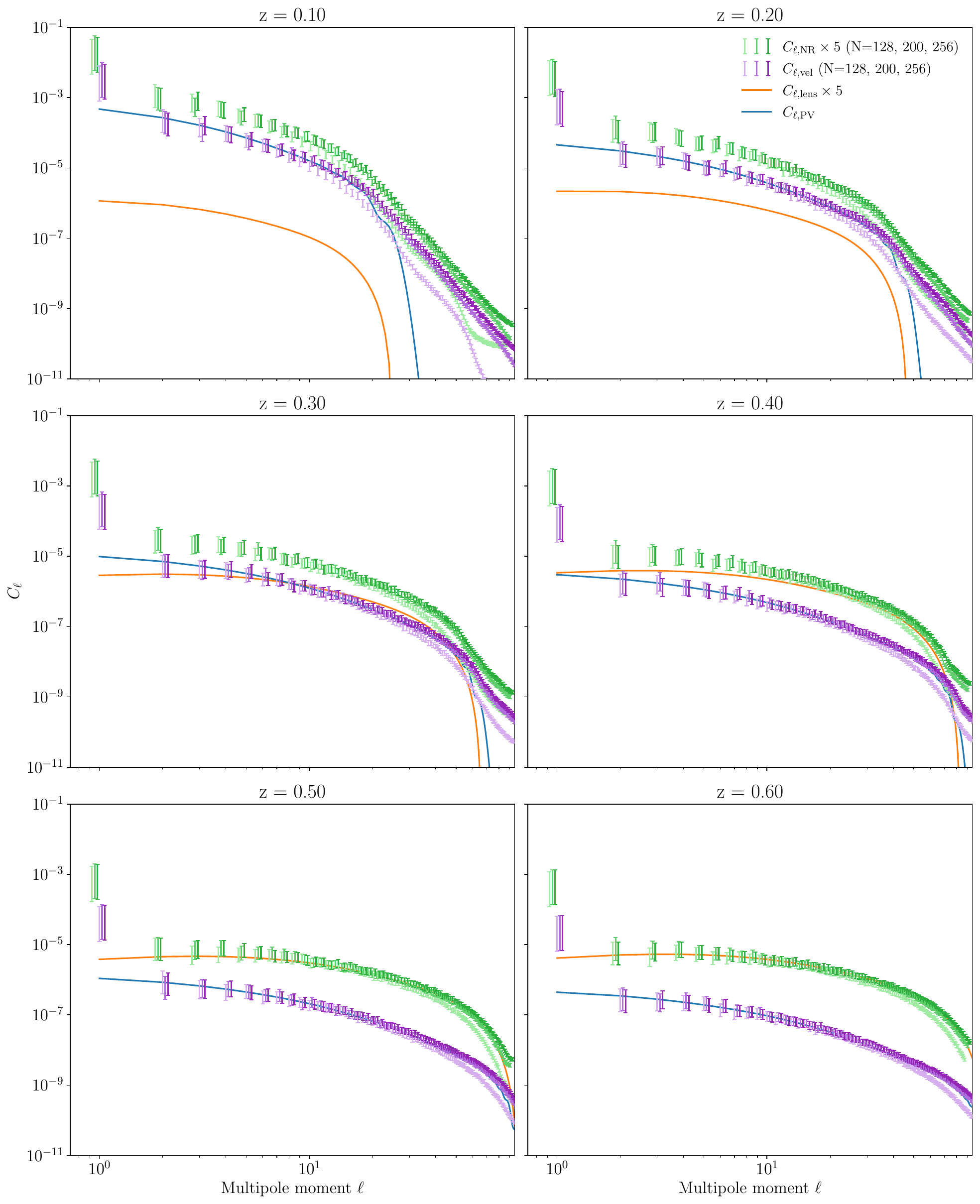}
    \caption{Angular power spectra of magnitude fluctuations $C_\ell^{\delta m \, \delta m}$ at redshifts $z = 0.1$ to $z = 0.6$ across three simulation resolutions $N = \{ 128, 200, 256 \}$. The green points show the NR measurement $C_{\ell,{\rm NR}}$ with the lightest green representing the coarsest resolution ($N=128$) and the green increasing in darkness as the resolution increases. The purple points show the velocity-only measurement $C_{\ell,{\rm vel}}$ with the lightest purple representing the coarsest resolution ($N=128$) and the purple increasing in darkness as the resolution increases. Error bars represent the standard deviation across 20 observer positions. The $C_{\ell,{\rm NR}}$ have been scaled by a factor of 5 to improved the ability to distinguish the cases. $C_{\ell,{\rm NR}}$ and $C_{\ell,{\rm vel}}$ are converged at $N = 256$.}
    \label{fig:clallres}
\end{figure*}

\subsection{Growth rate offset}
\label{sec:results_f}

If we neglect the effects of gravitational lensing when fitting magnitude correlations with velocity-only models, we may obtain a bias in the inferred growth rate.  We used the Fisher matrix method described in Sec.~\ref{sec:fisher} to forecast the resulting offset in growth rate, $\Delta f$, using the observer mean NR and velocity-only angular power spectra as inputs to the calculation.  We compare this offset with a forecast of the statistical error $\sigma_f$ arising from cosmic variance and measurement noise for an ideal all-sky survey described in Sec.~\ref{sec:fisher}.  In Fig.~\ref{fig:sysoff256}, we present, as a percentage of the fiducial growth rate, the systematic offset compared to the statistical error, as a function of redshift.

At $z \leq 0.2$ the statistical uncertainty is $3.1$--$4.1\%$ of the fiducial growth rate, while the systematic offset is $< 1\%$. This reflects the angular power spectrum results of Section~\ref{sec:results_cl} where at low redshift, the full NR and velocity-only measurements are in close agreement, and the resulting bias in $f$ is negligible.

As redshift increases, the systematic offset due to neglecting the lensing effect grows rapidly, reaching $7.5\%$ at $z = 0.3$, $38.9\%$ at $z = 0.4$, and exceeding $100\%$ by $z = 0.5$. The statistical error increases more gradually, from $3.1\%$ at $z = 0.1$ to $11\%$ at $z = 0.4$. This indicates that the standard velocity-only modelling would lead to significantly biased growth rate constraints if applied to data at redshifts $z \ge 0.3$ for this survey configuration, without accounting for these additional effects.

\subsection{Convergence tests}

We repeated our measurements across the three simulation resolutions $N = \{ 128, 200, 256 \}$ to test for numerical convergence of the results.  In Fig.~\ref{fig:clallres} we show the full NR and velocity-only angular power spectra measurements for the three resolutions, demonstrating convergence by $N = 256$. On the smallest scales there is some non-convergence, but this is below the initial data power cutoff. As we focus on larger scales $k < 0.09 \, h^{-1}$\unit{Mpc}), we see convergence for all relevant scales above the initial cutoff. The average difference between $N = 256$ and $N=200$ is $17.8\%$ for all $\ell > 1$, indicating that the simulation is sufficiently converged at $N = 256$, and the detected difference in power between $C_{\ell,{\rm NR}}$ and $C_{\ell,{\rm vel}}$ for a given simulation resolution is not due to the finite difference error of the simulation.

\begin{figure*}
    \centering
    \includegraphics[width=\textwidth]{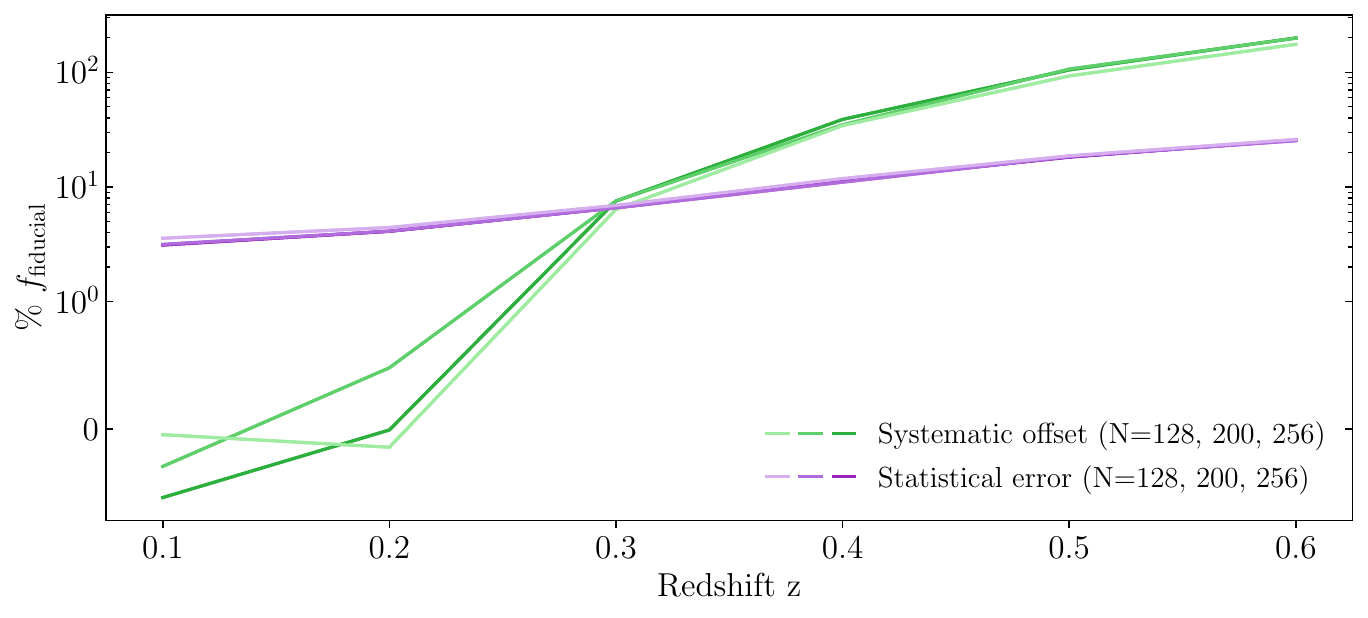}
    \caption{Convergence test of the systematic offset $\Delta f$ (green) and statistical error $\sigma_f$ (purple) across three simulation resolutions $N = \{ 128, 200, 256 \}$. The systematic offset and the statistical error are converged at $N = 256$, demonstrating that the measured systematic offset is robust to the simulation resolution.}
    \label{fig:sysoffallres}
\end{figure*}

In Fig.~\ref{fig:sysoffallres} we show a convergence test of the systematic offset $\Delta f$ (green) and statistical error $\sigma_f$ (purple) for the three simulation resolutions. The systematic offset and the statistical error are converged at $N = 256$, demonstrating that the inferred growth rate offset is robust to the simulation resolution.

\section{Discussion}
\label{sec:discussion}

Current PV surveys \citep[e.g.][]{2023MNRAS.525.1106S, 2023Msngr.190...46T, 2023MNRAS.519.4589C, 2025A&A...694A...1R} are restricted to a redshift range $z \lesssim 0.1$--$0.2$, due to increasing uncertainty in distance measurements at higher redshifts.  Within this redshift range, we find that the systematic offset in the inferred growth rate due to neglecting lensing and inhomogeneous curvature effects is sub-dominant to the statistical error, indicating that velocity-only FLRW modelling is an accurate approximation for growth rate inference at current levels of statistical precision.

However, for future surveys probing higher redshifts, such as those enabled by improved standard candle calibrations or next-generation spectroscopic surveys, we find that the contribution from non-velocity effects to magnitude fluctuations becomes more significant. These results support previous analyses performed within a perturbed FLRW framework \citep[e.g.,][]{2017MNRAS.467..259M}.  At $z \gtrsim 0.3$, we find that the systematic growth rate offset due to neglecting the lensing contribution exceeds the statistical error forecast for a baseline future survey configuration. At $z = 0.4$, lensing and peculiar velocity contributions to $C_{\ell}^{\delta m, \delta m}$ are of the same order. Therefore, cosmological inference at these redshifts ignoring lensing contributions may systematically overestimate $f$ by $\mathcal{O}(f)$.

We note that the statistical errors presented here represent a lower bound on the true uncertainty, as our Fisher matrix forecast varies only a single parameter, the growth rate of structure.  Additional model parameters, such as non-linear velocity dispersion or velocity bias \citep{2014MNRAS.445.4267K, 2026A&A...708A.219Q}, would need to be simultaneously constrained for more realistic analyses. Since these non-linear effects in PV analysis are suppressed by a factor of $k^{-2}$ in the PV kernel, we expect that marginalising over these additional parameters would only contribute to a small increase in the statistical error in $f$. Although this would be relatively marginal, any increase in the statistical error in $f$ would proportionately render the systematic offset less significant.

The NR simulations used in this analysis contain no particles, so we are unable to simulate the formation of collapsed structures such as dark matter halos. As a result, the density field is resolved only on scales above the grid resolution, and effects associated with non-linear small-scale structure, such as galaxy bias and non-linear velocity dispersion, are not captured. A higher-resolution simulation that resolves halo formation would enable a more complete treatment of the relationship between the observed galaxy distribution and the underlying matter field, as incorporated by \citet{2019PhRvD..99b3527G} and \citet{2019PhRvD.100j3533E}, and is in progress for the \texttt{Einstein Toolkit} by \citet{2023PhRvD.108j3534M}.

Our analysis is restricted to the angular power spectrum of magnitude fluctuations on two-dimensional redshift slices. An extension to fully three-dimensional density and velocity fields would enable complementary analyses, including cross-correlations between the density and velocity fields and the study of redshift-space distortions.

%

\section{Conclusions}
\label{sec:conclusion}

Using numerical relativity simulations, we have quantified the impact of peculiar velocities, gravitational lensing and inhomogeneous curvature in an inhomogeneous spacetime on growth rate measurements inferred from magnitude fluctuations. The FLRW framework which underpins standard PV analyses interprets magnitude fluctuations as arising solely from the Doppler effect of source and observer motion. In a fully General Relativistic treatment, the propagation of light through a varying gravitational potential introduces additional contributions from inhomogeneous curvature and gravitational lensing.  We studied whether neglecting these contributions in an FLRW model produces a significant systematic bias in the inferred growth rate of structure. We do this by measuring the angular power spectrum of magnitude fluctuations $C_\ell^{\delta m \, \delta m}$, and comparing results for the full NR power spectrum, $C_{\ell,{\rm NR}}$, with a velocity-only measurement, $C_{\ell,{\rm vel}}$, computed from the simulated velocity field alone.

Our key findings are as follows:
\begin{itemize}
    \item At $z \lesssim 0.2$, $C_{\ell,{\rm NR}}$ and $C_{\ell,{\rm vel}}$ are in close agreement across the full range of multipoles, confirming that magnitude fluctuations in this regime are dominated by the Doppler contribution from peculiar velocities. The corresponding systematic offset in the growth rate is sub-dominant ($< 1\%$) to the statistical fluctuations of a future survey configuration.
    \item At $z = 0.3$, an excess in $C_{\ell,{\rm NR}}$ relative to $C_{\ell,{\rm vel}}$ becomes apparent, continuing to grow with increasing redshift. This excess reflects the increasing contribution of inhomogeneous curvature and gravitational lensing along longer path lengths through the inhomogeneous spacetime.
    \item The systematic offset in the growth rate grows rapidly with redshift, reaching ${\sim} 8\%$ at $z = 0.3$, ${\sim} 39\%$ at $z = 0.4$, and exceeding $100\%$ by $z = 0.5$, while the statistical error grows more gradually. Beyond $z=0.3$ curvature and lensing effects would dominate over statistical uncertainties in growth rate measurements.
    \item We demonstrated convergence as a function of simulation resolution of the angular power spectrum measurements and the systematic offset in the growth rate.
\end{itemize}

Our results imply that for current PV surveys, which extend to $z \lesssim 0.1$--$0.2$, neglecting gravitational lensing and inhomogeneous curvature does not significantly bias growth rate inference at present levels of statistical precision. Future surveys pushing growth rate measurements to intermediate redshifts $z\approx 0.3$ will require appropriate modelling of non-velocity effects on magnitude fluctuations in order to reach percent level of accuracy.

%

\section*{Acknowledgements}

We are grateful to Ryan Turner and Leonardo Giani for useful comments during the completion of this project.  This research was conducted by the Australian Research Council Centre of Excellence for Gravitational Wave Discovery (project number CE230100016) and funded by the Australian Government.  AN would like to acknowledge the financial support received through the award of a Research Training Program Stipend scholarship by Swinburne University. 
Support for HJM was provided by NASA through the NASA Hubble Fellowship grant HST-HF2-51514.001-A awarded by the Space Telescope Science Institute, which is operated by the Association of Universities for Research in Astronomy, Inc., for NASA, under contract NAS5-26555. HJM was also supported by the Kavli Institute for Cosmological Physics through an endowment from the Kavli foundation and its founder Fred Kavli. This work was performed on the OzSTAR national facility at Swinburne University of Technology. The OzSTAR program receives funding in part from the Astronomy National Collaborative Research Infrastructure Strategy (NCRIS) allocation provided by the Australian Government, and from the Victorian Higher Education State Investment Fund (VHESIF) provided by the Victorian Government. Some of the results in this paper have been derived using the \texttt{HEALPix} \citep{Gorski:2005} package. Simulations and post-processing analyses used in this work were performed with resources provided by the University of Chicago's Research Computing Center. 

\section*{Data availability}

The data underlying this article will be shared on reasonable request to the corresponding author.

\bibliographystyle{mnras}
\bibliography{main}

\appendix

\section{Angular power spectrum models for the density and radial velocity fields}
\label{sec:derivation}

In this section, we derive the model for the angular power spectrum of radial velocities given in Eq.~\eqref{eq:cluu} and Eq.~\eqref{eq:wlu}.  We define the three-dimensional matter overdensity field at position $\mathbf{x}$ as $\delta(\mathbf{x})$, and the galaxy overdensity assuming linear bias $b$ as $\delta_g(\mathbf{x}) = b \, \delta(\mathbf{x})$.  The projected galaxy density in real space at angular position $\hat{\mathbf{x}} = (\theta,\phi)$, which we write as $\Delta(\hat{\mathbf{x}})$, is then given by integrating $\delta_g(\mathbf{x})$ over the radial coordinate $\chi$,
\begin{equation}
\Delta(\hat{\mathbf{x}}) = b \int d\chi \, w(\chi) \, \delta(\mathbf{x}) ,
\end{equation}
where $w(\chi)$ is the projection (weight) function, which is determined by the redshift distribution of galaxies, normalized such that $\int d\chi \, w(\chi) = 1$.  Expressing this equation in terms of the Fourier amplitudes of the density field, $\tilde{\delta}(\mathbf{k})$,
\begin{equation}
\Delta(\hat{\mathbf{x}}) = \frac{b}{(2\pi)^3} \int d\chi \, w(\chi) \int d^3\mathbf{k} \, \tilde{\delta}(\mathbf{k}) \, e^{-i \mathbf{k}.\mathbf{x}} .
\end{equation}
Using the spherical harmonic decomposition of a plane wave,
\begin{equation}
e^{i \mathbf{k}.\mathbf{x}} = 4 \pi \sum_{\ell m} i^\ell \, j_\ell(k\chi) \,
Y_{\ell m}(\hat{\mathbf{x}}) \, Y_{\ell m}^*(\hat{\mathbf{k}}) ,
\end{equation}
where $j_\ell$ is the spherical Bessel function and $Y_{\ell m}$ are the spherical harmonics, we can express the projected density in spherical harmonics, $\Delta(\hat{\mathbf{x}}) = \sum_{\ell m} \Delta_{\ell m} Y_{\ell m}(\hat{\mathbf{x}}) = \sum_{\ell m} \Delta_{\ell m}^* Y_{\ell m}^*(\hat{\mathbf{x}})$, obtaining,
\begin{equation}
\Delta_{\ell m}^* = \frac{b \, (i^\ell)^*}{2\pi^2} \int d\chi \, w(\chi) \int d^3\mathbf{k} \, \tilde{\delta}(\mathbf{k}) \, j_\ell(k\chi) \, Y_{\ell m}(\hat{\mathbf{k}}) .
\end{equation}
We can re-order the integration such that
\begin{equation}
\Delta_{\ell m}^* = \frac{(i^\ell)^*}{2\pi^2} \int d^3\mathbf{k} \, \tilde{\delta}(\mathbf{k}) \, Y_{\ell m}(\hat{\mathbf{k}}) \, W_\ell(k) ,
\end{equation}
where,
\begin{equation}
W^\delta_\ell(k) = b \int d\chi \, w(\chi) \, j_\ell(k\chi) .
\end{equation}
The angular power spectrum is defined by $C_\ell = \langle \Delta_{\ell m} \, \Delta_{\ell m}^* \rangle$.  If we use the definition of the power spectrum $\langle \tilde{\delta}(\mathbf{k}) \delta(\mathbf{k}') \rangle = (2\pi)^3 \, P(k) \, \delta_D(\mathbf{k}-\mathbf{k}')$, where $\delta_D$ is the Dirac delta function, and the normalization $\int d^2\hat{\mathbf{k}} \, |Y_{\ell m}(\hat{\mathbf{k}})|^2 = 1$, we obtain,
\begin{equation}
\langle \Delta_{\ell m} \Delta_{\ell m}^* \rangle = \frac{2}{\pi} \int dk \, k^2 \, P(k) \, \left[ W^\delta_\ell(k) \right]^2 .
\end{equation}
We now obtain the analogous result for the projected radial velocity.  In linear theory we have the continuity equation:
\begin{equation}
\tilde{\mathbf{v}}(\mathbf{k}) = - i H f \frac{\mathbf{k}}{k^2} \tilde{\delta}(\mathbf{k}) .
\end{equation}
The radial velocity $u$ at vector $\mathbf{x} = r\hat{\mathbf{x}}$ from the observer is defined by $u(\mathbf{x}) = \mathbf{v}(\mathbf{x}) . \hat{\mathbf{x}}$.  By substituting in the Fourier transform $\tilde{\mathbf{v}}(\mathbf{k})$ and the continuity equation, we obtain
\begin{equation}
u(\mathbf{x}) = \frac{- i H f}{(2\pi)^3} \int d^3\mathbf{k} \, \frac{\tilde{\delta}(\mathbf{k})}{k} \, (\hat{\mathbf{x}}.\hat{\mathbf{k}}) \, e^{-i \mathbf{k}.\mathbf{x}} .
\end{equation}
Noting that $(\hat{\mathbf{x}}.\hat{\mathbf{k}}) \, e^{-i \mathbf{k}.\mathbf{x}} = i \frac{d}{d(k\chi)} \left[ e^{-i \mathbf{k}.\mathbf{x}} \right]$ and expressing $e^{-i \mathbf{k}.\mathbf{x}}$ using the plane wave expansion, we obtain,
\begin{equation}
u(\mathbf{x}) = \frac{H f}{2\pi^2} \sum_{\ell m} (i^\ell)^* \int d^3\mathbf{k} \, \frac{\tilde{\delta}(\mathbf{k})}{k} \, j_\ell'(k\chi) \, Y_{\ell m}(\hat{\mathbf{k}}) \, Y_{\ell m}(\hat{\mathbf{x}}) ,
\end{equation}
where $j_\ell'$ is the derivative of the spherical Bessel function.   In a spherical harmonics expansion $u(\mathbf{x}) = \sum_{\ell m} u_{\ell m}(\chi) Y_{\ell m}(\hat{\mathbf{x}})$ we can therefore identify,
\begin{equation}
u_{\ell m}^*(\chi) = \frac{H f (i^\ell)^*}{2\pi^2} \int d^3\mathbf{k} \, \frac{\tilde{\delta}(\mathbf{k})}{k} \, j_\ell'(k\chi) \, Y_{\ell m}(\hat{\mathbf{k}}) .
\end{equation}
The projected radial velocity is,
\begin{equation}
U(\hat{\mathbf{x}}) = \int d\chi \, w(\chi) \, u(\mathbf{x}) ,
\end{equation}
and hence we can write the spherical harmonic coefficients,
\begin{equation}
U_{\ell m}^* = \frac{(i^\ell)^*}{2\pi^2} \int d^3\mathbf{k} \, \tilde{\delta}(\mathbf{k}) \, Y_{\ell m}(\hat{\mathbf{k}}) \, W^v_\ell(k) ,
\end{equation}
where,
\begin{equation}
W^v_\ell(k) = \frac{H f}{k} \int d\chi \, w(\chi) \, j_\ell'(k\chi) .
\end{equation}
We can now use the recurrence relation $(2\ell + 1) j_\ell'(x) = \ell \, j_{\ell-1} - (\ell + 1) j_{\ell+1}$ to write,
\begin{equation}
(2\ell + 1) W^v_\ell(k) = \frac{H \beta}{k} \left[ \ell \, W^\delta_{\ell-1}(k) - (\ell + 1) W^\delta_{\ell+1}(k) \right] ,
\end{equation}
where $\beta = f/b$, as given in Eq.~\eqref{eq:wlu}.  Proceeding in a similar manner to above, we can write down the radial velocity angular power spectrum as given in Eq.~\eqref{eq:cluu}:
\begin{equation}
< U_{\ell m} U_{\ell m}^* > = \frac{2}{\pi} \int dk \, k^2 \, P(k) \, [W^v_\ell(k)]^2 .
\end{equation}

\end{document}